\documentclass{aa}

\usepackage{graphicx}
\usepackage{txfonts}
\usepackage{natbib}

\newcommand{\kms}{km~s$^{-1}$}

\newcommand{\rsun}{$R_{\odot}$}
\newcommand{\msun}{$M_{\odot}$}
\newcommand{\erf}{\rm erf}

\begin{document}

\title{Possible evidence of induced repetitive magnetic reconnection \\ in a superflare from a young solar-type star}

\author{S. Mancuso\inst{1}\and D. Barghini\inst{1,2}\and D. Telloni\inst{1}}

\institute{
INAF -- Osservatorio Astrofisico di Torino, via Osservatorio 20, 10025 Pino Torinese, Italy \and
Dipartimento di Fisica, Universit\`a degli Studi di Torino, via Pietro Giuria 1, 10125 Torino, Italy\\ \email{salvatore.mancuso@inaf.it}}
\date{Received / Accepted}

\abstract{
We report the detection of multiple quasi-periodic pulsations (QPPs) observed during the flaring activity of KIC 8414845, a young, active solar-type star observed by the Kepler mission launched by NASA. 
We analyzed the QQP signal using a data-driven, nonparametric method called singular spectrum analysis (SSA), which has never been utilized previously for analyzing solar or stellar QPPs.
Because it is not based on a prescribed choice of basis functions, SSA is particularly suitable for analyzing nonstationary, nonlinear signals such as those observed in QPPs during major flares. 
The analysis has revealed that the apparent anharmonic shape of the QPP in this superflare results from a superposition of two intrinsic modes of periods of 49 min and 86 min, which display quasi-harmonic behaviors and different modulation patterns.
The two reconstructed signals are consistent with slow-mode transverse and/or longitudinal magnetohydrodynamic  oscillations excited in a coronal loop inducing periodic releases of flaring energy in a nearby loop through a mechanism of repetitive reconnection.
The peculiar amplitude modulation of the two modes evinced by SSA favors the interpretation of the observed QPP pattern as due to the excitation in a coronal loop of the second harmonic of a standing slow-mode magnetoacoustic oscillation and a global kink oscillation periodically triggering magnetic reconnection in a nearby loop. 
Concurrent interpretations cannot however be ruled out on the basis of the available data.

}
\titlerunning{Possible evidence of induced repetitive magnetic reconnection in a stellar superflare}
\authorrunning{Mancuso et al.}
\maketitle

\begin{figure*}
\centering
\includegraphics[width=18cm]{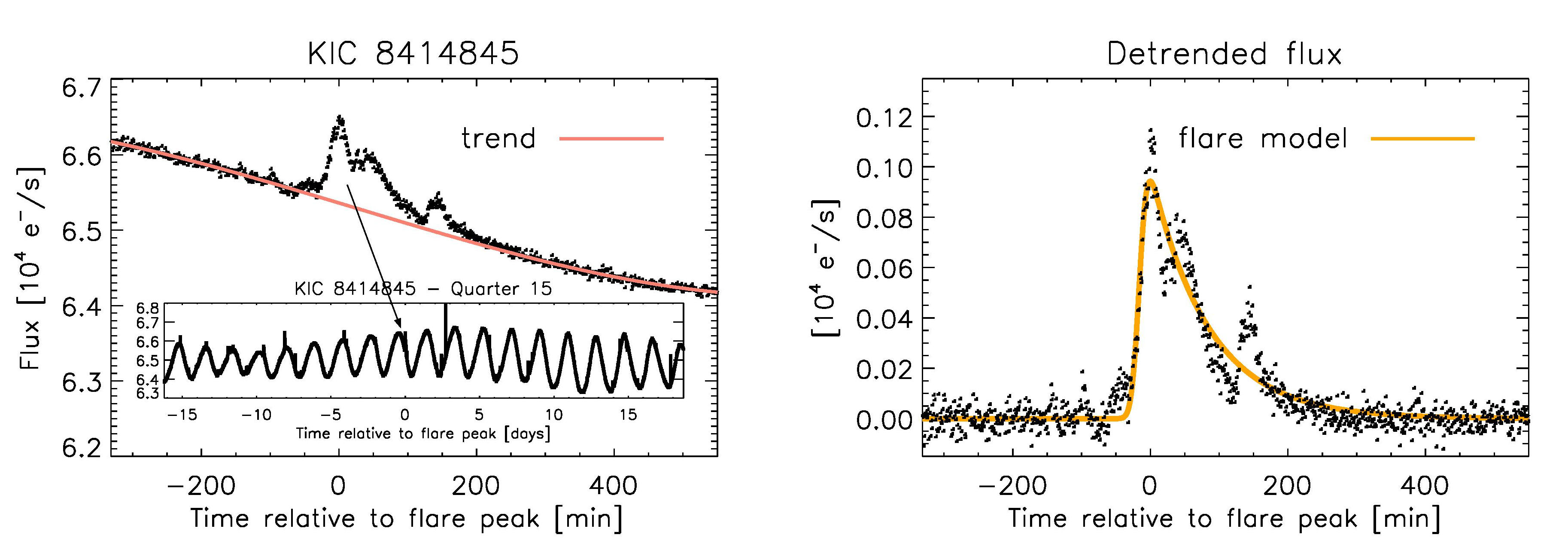}
\caption{{\it Left panel}: Light curve showing the flare on KIC 8414845 with start time at the flare peak (MJD 56285.43). A third-order polynomial function was fitted to the background to account for the modulation of the light curve due to the binary nature of the star (see inset). {\it Right panel}: The flare model by \cite{gryciuk2017} superposed on the detrended light curve of the flare.}
\label{Fig1}
\end{figure*}

\section{Introduction}

The analysis of white-light data collected by the Kepler mission launched by NASA indicates that extremely energetic flare activity occurs on solar-type stars (\citealt{Maehara2012}).
The presence of superflares (flares with energy $>10^{33}$ erg) on solar-like stars opens the question of the possible occurrence of such catastrophic phenomena on our Sun; these superflares might be capable of affecting the chemistry of Earth's atmosphere or even lead to extinction events.
Light curves of superflares often display pronounced oscillatory patterns of emission, known as quasi-periodic pulsations (QPPs), with characteristic periods of tens of minutes (e.g., \citealt{Balona2015}; \citealt{Pugh2016}).
The specific physical explanation for their generation is still under debate although a variety of mechanisms have been proposed in the literature.
The most viable mechanisms are generally divided into two groups: those in which the flare emission is modulated by magnetohydrodynamic (MHD) oscillations and those based on some regime of repetitive magnetic reconnection (\citealt{McLaughlin2018}). 
A main issue with QPPs is to distinguish the observed variability with that associated with stochastic processes that manifest themselves as colored noise (\citealt{Pugh2017}).
Only a few percent of the superflares observed by Kepler show statistically significant QPPs (\citealt{Pugh2016}). 
Indeed, events that have been previously considered as containing signatures of QPPs can be  adequately described via a power-law model of the Fourier power spectrum (\citealt{Inglis2015}).
This problem is even more critical whenever multiple QPPs are detected in the same event.
To our knowledge, only one clear case (\citealt{Pugh2015}) has been reported of a white-light superflare displaying multiple QPPs.
Such events are however of paramount importance in that they can provide additional constraints on the interpretation and understanding of the physical processes occurring in superflares near the stellar surface while stimulating interesting questions about the theoretical mechanisms that may be responsible for their production.

In this work, we apply for the first time a powerful alternative technique, called singular spectrum analysis (SSA; \citealt{Broomhead1986}), to analyze multiple QPPs detected in the light curve of a superflare that occurred in KIC 8414845, a rapidly rotating young, solar-type star observed by the Kepler mission by NASA.
In constrast to Fourier-based techniques, SSA tries to overcome the problems of finite sample length and noisiness of time series using a data-adaptive basis set, that is, a decomposition basis locally defined starting from the signal itself.
Although SSA has been already successfully used to analyze solar coronal data (\citealt{Mancuso2015,Mancuso2016,Mancuso2018}), this is the first time this technique is applied to the analysis of QPPs occurring in flares.
We show that SSA is particularly suited to study nonlinear oscillations such as the QPPs occurring in superflares; this allows for a detailed determination of these essentially anharmonic signals, while simultaneously providing a robust tool to investigate the statistical significance of the signals against colored noise.
The reconstruction of the instantaneous amplitude and period of the two oscillations provided by SSA will finally yield a means to delve into the possible physical mechanisms underlying the observed modulation of the optical radiation emitted during this exceptional event.  
This paper is structured as follows. Section 2 describes the properties of the star hosting the superflare analyzed in this work. Section 3 presents the data analysis of the Kepler data and the results obtained by applying the SSA technique outlined in the Appendix to the superflare profile. Section 4 contains the discussion and interpretation of the results obtained with the SSA analysis. Finally, we summarize our findings in Section 5.

\section{Kepler observations} 

The majority of photometric observations of the target stars collected by the Kepler mission were made in long-cadence mode (one data point about each half an hour). 
However, a few thousand of these have been observed with 1 min short cadence exposures for several days, thus allowing QPPs in flares to be detected.
The superflare analyzed in this paper occurred on KIC 8414845 (2MASS J18575814+4425429) and its light curve (Fig. 1) can be found in the 1 min short-cadence observations collected by the Kepler mission  from Quarter 15, starting on 2012 December 7 and ending on 2013 January 11. 
Its energy was estimated at $5.7^{+4.3}_{-2.5}\cdot10^{34}$ erg (\citealt{Pugh2016}), with an enhancement of the flux of about 0.1\%.
The data was retrieved through the onlne Mikulski Archive for Space Telescopes (MAST) database\footnote{https://archive.stsci.edu/.}.
According to the data stored in the online GAIA second data release (DR2)\footnote{http://gea.esac.esa.int/archive/.}, KIC 8414845 (identified as Gaia DR2 2106705789681987840) has a parallax of $1.643 \pm 0.016$ mas, corresponding to a distance of $602.8-614.6$ pc.
The physical properties, retrieved from the revised stellar catalog of \cite{Mathur2017}, are the following: effective temperature $T_{\rm eff} = 5693 \pm 154$ K, log surface gravity $= 4.436^{+0.120}_{-0.165}$cm s$^{-2}$, metallicity (Fe/H) $= -0.400 \pm 0.300$, radius $R_\star = 0.899^{+0.224}_{-0.131}$ \rsun, and mass $M_\star = 0.804^{+0.112}_{-0.056}$ \msun. 
A rotation period of $1.88 \pm 0.22$ days was determined by \cite{Pugh2016}.
Given its frequent flare activity and its fast rotation period,  such that the fast rotation drives the emergence of strong magnetic fields, this is most probably a young, active solar-type binary star. 

\begin{figure*}
\centering
\includegraphics[width=13cm]{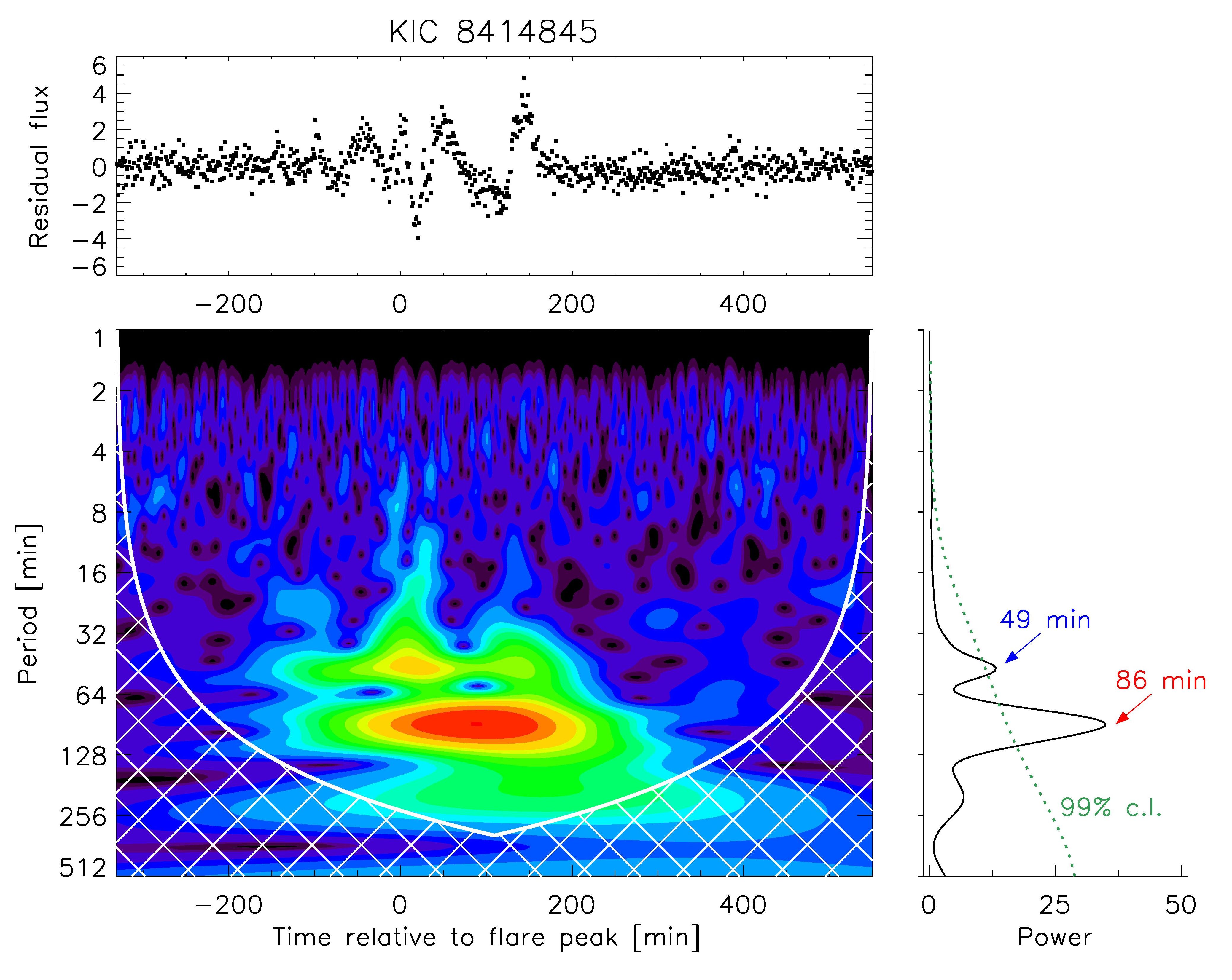}
\caption{Wavelet power spectrum of the residual light curve ({\it top panel}) after subtraction of the flare background and model. Red depicts the periods with greatest power. The hatched area denotes the cone of influence, that is, the region of the wavelet power spectrum in which edge effects become important. {\it Right panel}: Global power spectrum with 99\% c.l..}
\label{Fig2}
\end{figure*}

\section{Data analysis and results}

Two types of flux data, simple aperture photometry (SAP) and pre-search data conditioning simple aperture photometry (PDCSAP) in which instrumental effects are removed, are available for each light curve. 
In this work, we used the high-cadence ($\sim$ 1 min) SAP light curves because the PDCSAP post-processing could remove flare data that could be mistaken as outliers (\citealt{Balona2015}).
The QQP signal of the flare was obtained after detrending the light curve that is intrinsically contaminated by noise. 
In order to disentangle the flare trend and reduce the risk of introducing artificial signals that may be misinterpreted as QPPs, we applied an analytical model of the flare profile developed by \cite{gryciuk2017} given by
\begin{equation}
\scalebox{1.05}{
$f(t) = {\sqrt\pi\over 2}{\rm AC}\exp\left[{\rm D}({\rm B}-t)+{{\rm C}^2{\rm D}^2\over 4}\right]\left[\erf({Z})-\erf\left({Z}-{\it t}\over C\right)\right]$}
,\end{equation}
where $t$ is the time, A, B, C, and D are constants, ${\rm Z} \equiv {2{\rm B}+{\rm C}^2{\rm D}/(2{\rm C}),}$ and $\erf$ is the error function, defined as $\erf({\it t}) \equiv {2\over \sqrt{\pi}}\int_0^{\it t}\exp(-{\it s}^2)d{\it s}$.
This analytical time profile reflects the common action of two simultaneous processes, that is, energy deposition and energy dissipation by radiation and thermal conduction. 
The energy release process is assumed to have a Gaussian form and the energy dissipation rate an exponential form, where the reciprocal of the parameter $D$ represents the cooling time.
Figure 1 (right panel) shows the model superposed to the detrended light curve of the flare obtained after subtracting a third-order polynomial function to the background  to account for the local modulation of the light curve due to the binary nature of the star.
By removing the underlying flare profile, we thus emphasize any short-term variability. 

We first performed a wavelet analysis (\citealt{Torrence1998}) on the residual light curve, shown at the top of Fig. 2, using the Morlet wavelet, which provides a good balance between time and frequency localization.
The wavelet analysis indicates the presence of two significant peaks above the 99\% confidence level (c.l.) for red noise as clearly evinced from the global wavelet spectrum.
This result is consistent with the Lomb-Scargle periodogram analysis (\citealt{Scargle1982}) of the same detrended light curve for which two distinct peaks were found at 86.0 min and 49.0 min, both above the 99.9\% c.l. for red noise (see Fig. 3).
However, as already discussed, QPP patterns are essentially nonstationary oscillations, so that the results obtained with Fourier-based techniques are questionable. 
While the Fourier approach used in the Lomb-Scargle periodogram assumes that the signal has a harmonic shape with slowly modulated amplitude, the typical duration of QPP patterns is only a few oscillation cycles and their amplitude can be heavily modulated so that this intrinsic feature must be taken into proper account when investigating these events.
On the other hand, the outcome of wavelet analyses is highly sensitive to the choice of the assumed ad hoc mother function (\citealt{DeMoortel2000}; \citealt{Nakariakov2019}).
Finally, the standard procedure for significance testing in both the wavelet and Lomb-Scargle techniques is based on pure first-order autoregressive (AR1) red noise and, as such, it may not be appropriate when inspecting the power spectrum of the flare, which is expected to be power-law-like.
The clear anharmonicity and nonlinearity of the QPPs observed in the light curve of this stellar flare thus requires a proper self-adaptive technique such as SSA, which does not rely upon prescribed basis functions for their analysis and for which the statistical significance of the retrieved periodicities can be robustly inspected using proper noise modeling.

\begin{figure}[t!]
\centering
\includegraphics[width=9cm]{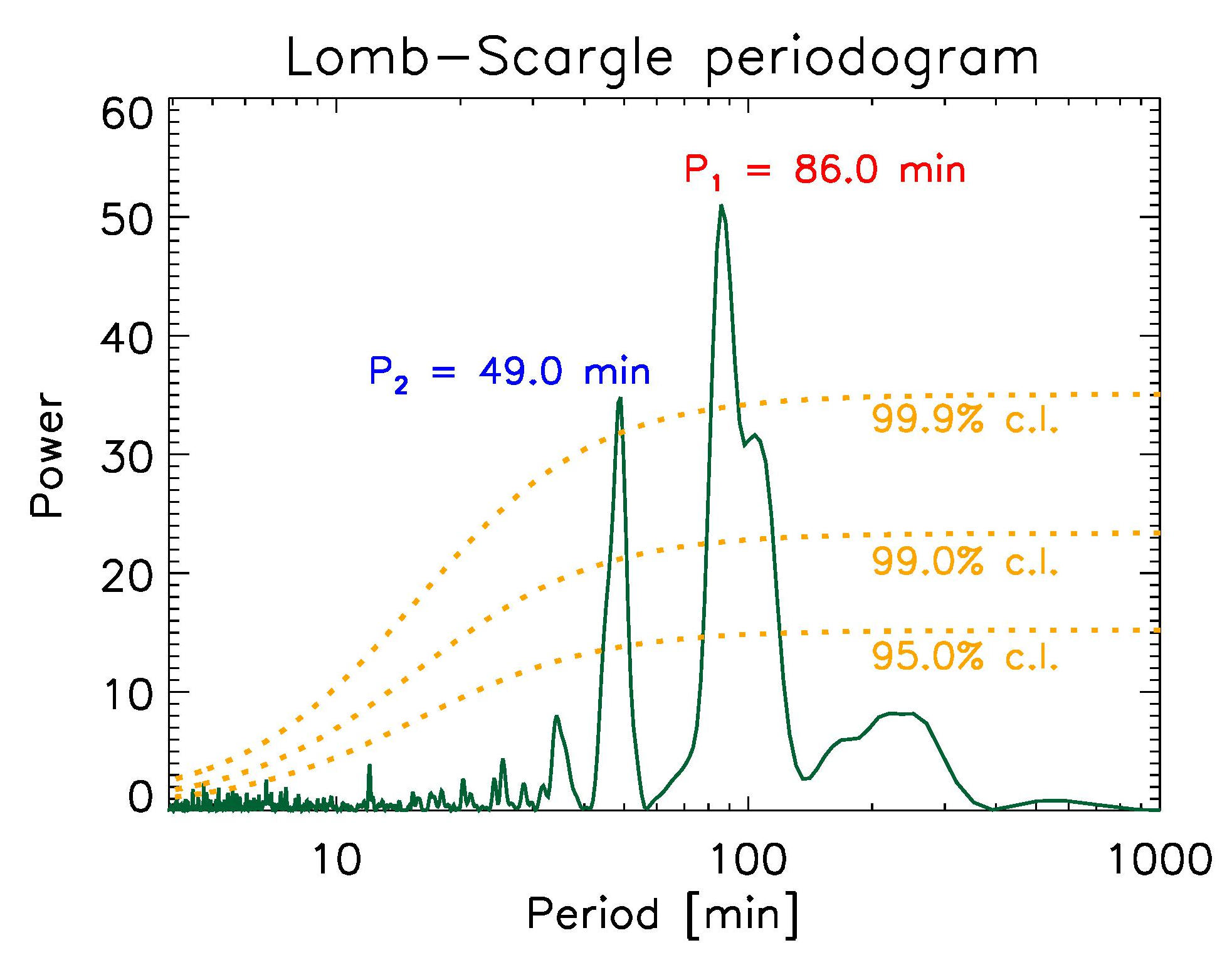}
\caption{Lomb-Scargle periodogram of the residual light curve removed from the flare background and model. The 95.0\%, 99.0\%, and 99.9\% c.l. for red noise are shown by a dotted line.}
\label{Fig3}
\end{figure}

Because it is not based on a prescribed choice of the basis functions, the SSA method is a nonparametric data-adaptive technique that is particularly suitable for analyzing nonstationary, nonlinear signals  (\citealt{Vautard1989}) such as those observed in QPPs during major flares. 
Since its use is uncommon in the astrophysical context and has never been applied in the analysis of flares or Kepler data, we briefly outline the theory involved in Appendix A. 
The timescales of the dynamics addressed by the SSA technique are bounded from above by the window width $M$, whose choice is given by a trade off between the amount of information that we wish to retain and the required degree of statistical significance.
Since SSA is typically successful at analyzing periods in the range ($M/5,M$), a window width of $M = 90$ was adopted in our analysis out of $N = 900$ data points, thus allowing us to capture oscillations between roughly 20 min and 90 min while still providing a high degree of statistical significance.
By arranging the singular values in decreasing order, it is qualitatively possible to distinguish the signal given by an initial steep slope, from the noise given by a ``flatter floor'' (see inset in Fig. 4).
A correct discrimination between the signal and colored noise of the SSA-derived oscillatory patterns can be obtained by applying a Monte Carlo analysis, which prevents the misinterpretation of stochastic oscillations due to colored noise as deterministic signals. 
For this event, a significance testing was conducted by comparing the data eigenvalues against those obtained by a large set of ideal surrogate time series generated with a Monte Carlo approach from power-law-like noise with the same total variance. 
The data and surrogate eigenvalues are therefore correlated via their dominant frequency (\citealt{Allen1996}, \citealt{Palus2004}).
Confidence intervals, based on the empirical distribution of the simulated data, were formed with 10000 surrogate realizations (see Fig. 4). 
We applied a stringent criterion such that the only data eigenvalues considered as significant were those lying outside the 99.9\% confidence interval of the corresponding eigenvalues ensemble of the surrogate realizations.

Oscillatory modes can be readily identified as pairs of nearly equal eigenvalues that have their principal components (PCs) with about the same period of oscillation and in quadrature with each other. 
This property allows us to properly isolate anharmonic oscillations with variable amplitudes from noisy data.
From visual inspection, two pairs of nearly equal singular values (PCs 1-2 and 3-4), each capturing an oscillatory mode, clearly stand out in Fig. 4 at frequencies around 0.01 and 0.02 min$^{-1}$. 
Both modes are highly significant, largely exceeding the 99.9\% c.l. in spectral power and thus very unlikely to be the result of colored noise.
The remaining singular values lie within the chosen c.l. and are thus considered as indistinguishable from stochastic noises.
The reconstructed components (RCs) of the two modes, RCs 1-2 (mode 1) and RCs 3-4 (mode 2), are shown in Fig. 5 and recover major portions of the variability in the original time series: the eigenvectors associated with these two pairs of RCs represent  more than half (55\%) of the total variance.
The Fourier spectrum of the two reconstructed modes yield similar periods as obtained in the Lomb-Scargle periodogram.

\begin{figure}[b]
\centering
\includegraphics[width=9cm]{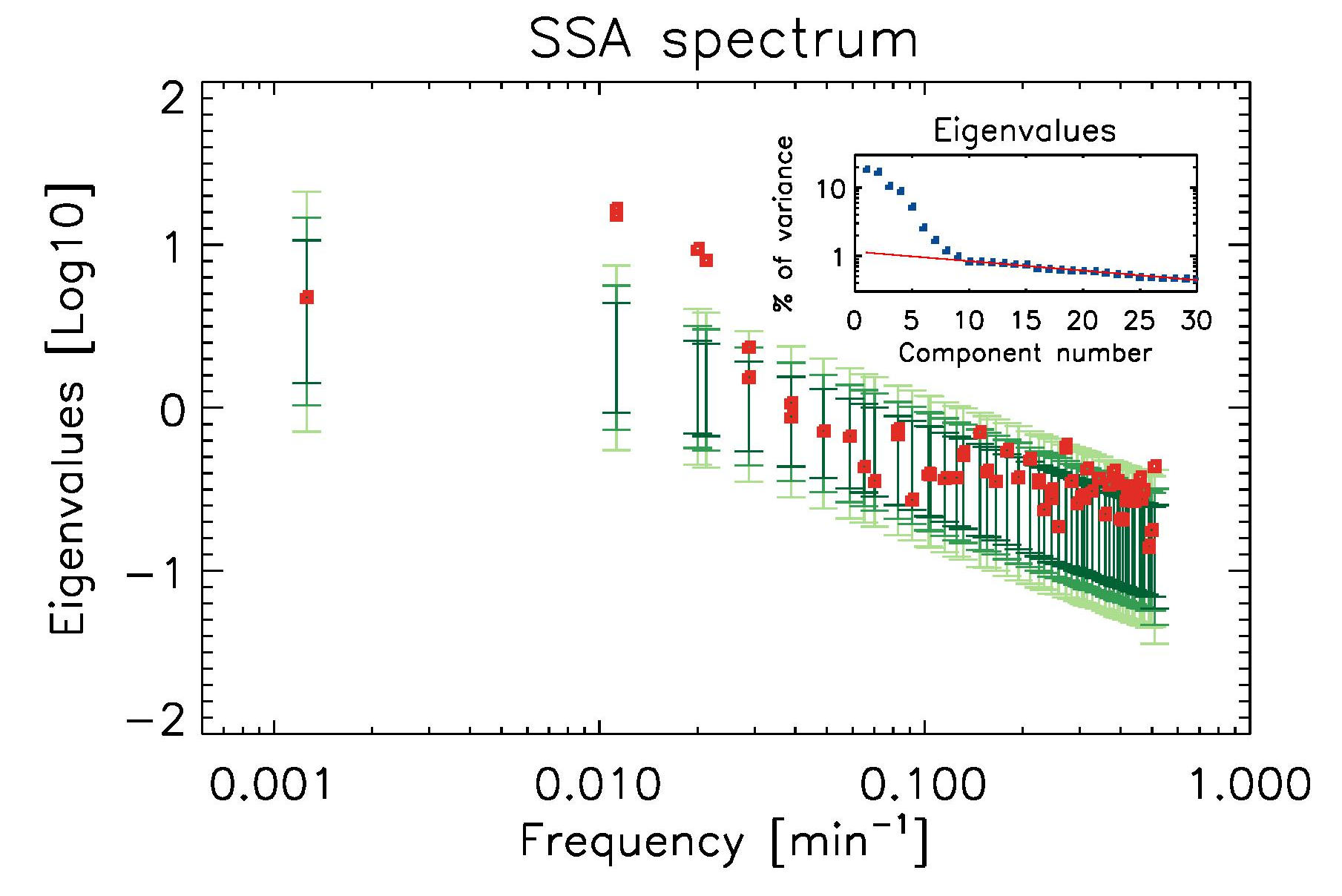}
\caption{Singular spectrum analysis spectrum of the detrended flare light curve with 95.0\%, 99.0\%, and 99.9\% confidence intervals (error bars in shades of green) computed from 10000 surrogate time series obtained via a Monte Carlo simulation.}
\label{Fig4}
\end{figure}

\section{Discussion and interpretation}

In flares, QPPs are usually attributed either to MHD oscillations in the flaring region or periodic modulation of the magnetic energy release region such as a reconnection site.
Intriguingly, the interaction of MHD waves with the reconnection site may lead to induced QPPs by means of a variety of proposed mechanisms (see a recent review by \citealt{McLaughlin2018}).
A question arises as to whether QPPs observed in white-light superflares are of the same nature as those directly observed in solar coronal structures.
Stellar flare light curves obtained by Kepler represent emission integrated in the optical passband from 400 nm to 900 nm.
In this wavelength range, the white-light continuum emitted by the dense superflare loops is mainly due to free-bound hydrogen recombination and hydrogen free-free processes (\citealt{Heinzel2018}).
White-light emission from solar loops has never been detected against the solar disk because typical electron densities in solar-flare loops do not significantly exceed $10^{12}$ cm$^{-3}$.
Moreover, solar white-light flares provide at most a contrast of only 0.01\% when integrated over the disk (\citealt{Haisch1991}), which is a figure that is about one or two order of magnitudes lower than what is typically observed in the stellar case (\citealt{Maehara2012}).
As for superflares of solar-type stars, the situation may differ since we expect much larger loop densities as the result of related strong evaporative processes.
The energy released by superflares is basically a part of the magnetic energy stored around very large starspots (\citealt{Notsu2013}) and so we also expect very high magnetic field strengths.
As in the case of the Sun, magnetic reconnection in coronal structures above starspots and downward transport to the lower atmospheric layers by electron beams and thermal conduction should be the most plausible scenario of energy release in superflares.
Based on a theoretical analysis, \cite{Heinzel2018} propose that because stellar flare loops have the possibility of being much larger and denser than solar loops, they can significantly contribute to the total superflare white-light emission; this is because stellar flare loops can be visible against the stellar disk and can occupy much larger areas than ribbons. 

\begin{figure}
\centering
\includegraphics[width=9cm]{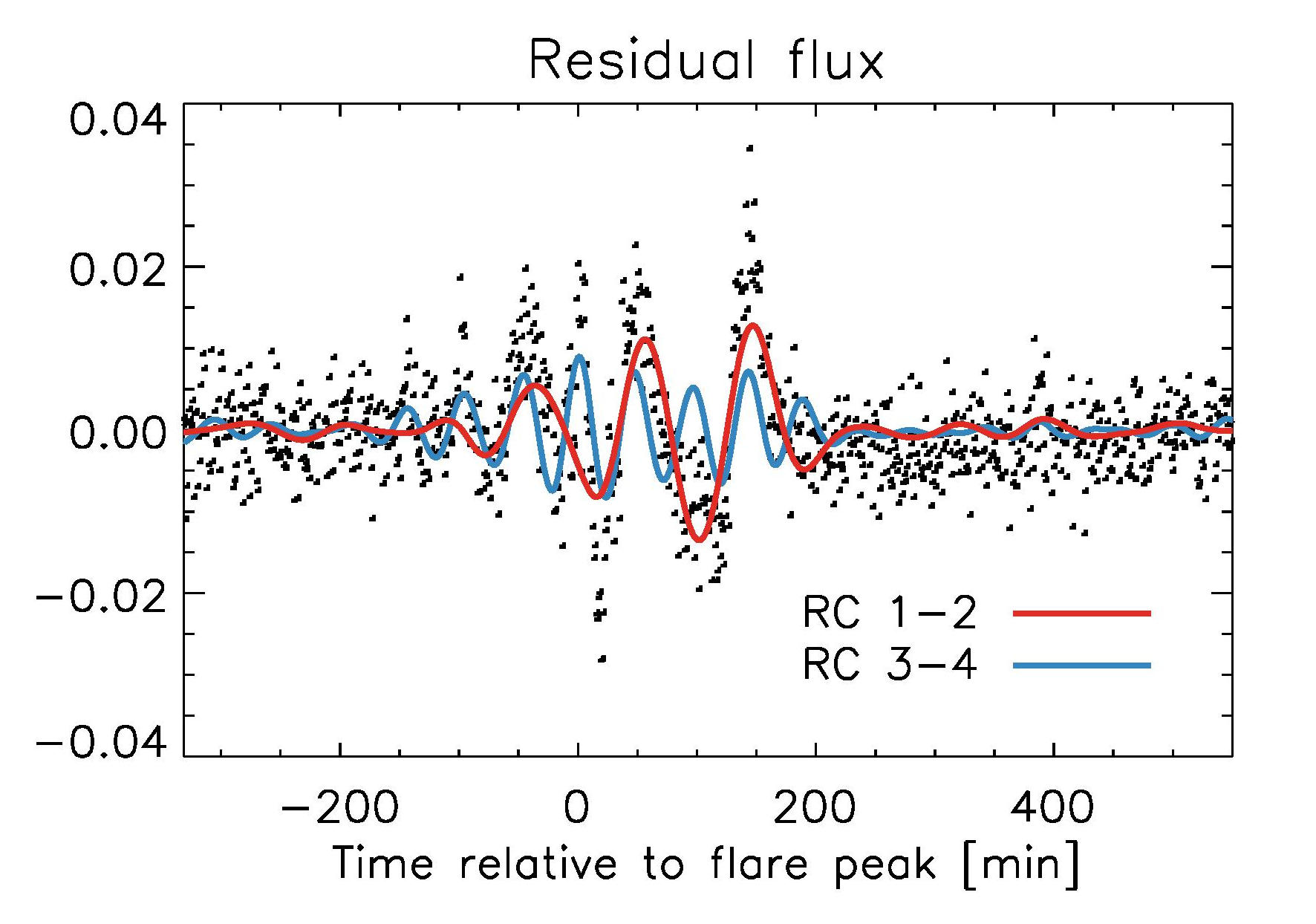}
\caption{First two couples of RCs,  RCs 1-2 (mode 1) and RCs 3-4 (mode 2), reproducing major portions of the variability in the original series of the detrended light curve of the flare.}
\label{Fig5}
\end{figure}

Different physical mechanisms have been proposed to explain oscillations of coronal plasma loops.
Several modes can be excited in loops, such as asymmetric (kink-type) modes, which are nearly incompressible and are manifested as transverse displacements, and symmetric (sausage-type) modes, which produce periodic contractions and expansions of the loops.
Sausage modes are the only modes that have significant density (and thus intensity) perturbations and are therefore often used to explain QPPs in solar flares.
In particular, long-period oscillations have classically been identified as slow-mode magnetoacoustic oscillations in flaring loops whose periods are determined by the ratio of the loop length and characteristic speed of the mode.
An impulsive energy release somewhere in the loop or near its footpoint, maybe induced by flaring activity, easily excites global longitudinal modes (\citealt{Nakariakov2004}; \citealt{Ofman2012}).
These oscillations typically show a rapid decay attributable to thermal conduction and compressive viscosity (\citealt{Wang2011}).
Unfortunately, without imaging observations, it is hard to conclusively claim what modes are responsible for the observed two periods. 
In the following discussion, we thus list a few alternative scenarios and attempt to select our favored scenario.  

A plausible scenario consistent with the observed coexistence of the two modes of periods $P_1 = 86$ min (mode 1) and $P_2 = 49$ min (mode 2) reconstructed in Figure 5 implicates the overlapping of the fundamental and harmonic modes of a standing slow magnetoacoustic wave excited in a stellar coronal loop.
Assuming that mode 1 corresponds to the fundamental mode, its period is determined by the ratio of the wavelength (double the length of the loop for the fundamental mode) to the phase speed (i.e., the tube speed) given by  $c_{\rm T} = c_{\rm S}v_{\rm A}/\sqrt{c_{\rm S}^2+v_{\rm A}^2} \approx c_{\rm S}$ in low-$\beta$ environments such as stellar coronae (\citealt{Roberts2000}). 
Typical temperatures $T = 10-30$ MK correspond to sound speeds $c_{\rm S} = 1.52\cdot10^4\sqrt{T}\approx 480-830$ \kms, thus resulting in loop lengths $L = P_1 c_{\rm T}/2 \approx 1.2-2.1\cdot 10^6$ km and major radii $L/\pi \approx 0.6-1.0$ \rsun. 
This is not surprising, given that the length of coronal loops in young stars have often been estimated as large as several times the stellar radius (e.g., \citealt{Favata2005}; \citealt{McCleary2011}; \citealt{Hartmann2016}; \citealt{Reale2018}).
In this scenario, the mode with period $P_2$ simply corresponds to the first harmonic of the standing slow magnetoacoustic wave excited in the loop.
The fact that the period of the harmonic is observed to be slightly larger than half of the fundamental ($P_2 \gtrsim P_1/2$) is  consistent with the theoretical analysis of slow-mode waves in a stratified atmosphere and the dispersive nature of the MHD waves (\citealt{Diaz2006}; \citealt{Inglis2009}).
Although viable, the above interpretation does not naturally explain the different amplitude modulation of the two oscillations and the fact that mode 2 appears before mode 1. In the following discussion, we thus investigate alternative scenarios involving the interaction between two loops.

Oscillatory phenomena in coronal loops can trigger and modulate the energy release during solar flares by means of several mechanisms linked to magnetic reconnection.
In particular, slow magnetoacoustic waves can periodically trigger magnetic reconnection by intermittently perturbing the plasma density in the vicinity of reconnection sites (\citealt{Chen2006}).
In this mechanism, density variations result in periodic variations of the electron drift speed and, as a consequence,  the periodic onset of anomalous resistivity that triggers the modulation of the energy release.
In such a scenario, the standing slow-mode wave in one of the loops modulates its interchange reconnection with the other loop (e.g., \citealt{Wang2015}).
A second possibility thus involves the excitement of two fundamental slow-mode waves with different periods ($P_1$ and $P_2$) in two different loops. 
In this interpretation, the flare was due to the interchange reconnection between the two loops.
The length of the second loop would be $L = P_2 c_{\rm T}/2 \approx 0.7-1.2\cdot 10^6$ km with a major radius on the order of $L/\pi \approx 0.3-0.6$ \rsun. 
Although viable, this latter interpretation does not naturally explain the observed nearly harmonic/fundamental ratio of the periods of the two reconstructed modes, which would appear somewhat coincidental.

In a third alternative scenario, our observations can be explained by assuming that two modes (a standing slow magnetoacoustic wave and a global kink oscillation) are excited in the same coronal loop; the latter  directly triggers the modulation of the emission of flaring energy in a nearby neutral point or loop (e.g., \citealt{Foullon2005}; \citealt{Nakariakov2006}; \citealt{Chen2006}).
In general, for loops with widths much smaller than the loop length $L$, two kink modes, slow and fast, are possible (e.g., \citealt{Roberts1984}; \citealt{Nakariakov2001}).
The period of global modes of fast kink oscillations is given by $P = 2L/c_{\rm K}$, where $c_{\rm K} \approx \sqrt{2/(1+\rho_e/\rho_0)}v_{\rm A}$ is the phase speed of the kink wave. 
Assuming that the loop length $L$ is constrained by the first harmonics of the standing slow magnetoacoustic wave excited in the same loop (mode 2), 
$P \approx 2L/(1.4v_{\rm A}) \approx 10 - 30$ min for $\rho_e/\rho_0 \ll 1$ and $\beta = 0.02-0.3$.
As for our observations, under the above assumption, the standing global kink fast magnetoacoustic mode has thus to be excluded from consideration for both modes 1 and 2.
Consequently, models such as those of \cite{Foullon2005} and \cite{Nakariakov2006}, which require the excitation of fast magnetoacoustic waves for inducing quasi-periodic episodes of reconnection, might be inappropriate for this event.
On the other hand, under the same assumption for the loop length $L$, the period of the global kink slow magnetoacoustic mode of the oscillating loop would be $P = 2L/c_{\rm K}\approx 98$ min, where $c_{\rm K} \approx c_{\rm T}$ (\citealt{Nakariakov2001}).  
This period is of the same order (albeit somewhat higher) as that estimated for mode 1. 
We should however take into account that the dispersion relation of standing slow kink modes  implies, under coronal conditions, that $c_{\rm T} < c_{\rm K} < c_{\rm s}$ (e.g., \citealt{Edwin1983}), so that the previous estimate must be rather intended as an upper limit (i.e., $c_{\rm K} \rightarrow c_{\rm T}$ only for $ka \rightarrow 0$, where $a$ is the radius of the tube).
Moreover, as already discussed, this is consistent with the theoretical analysis of slow-mode waves in a stratified atmosphere and the dispersive nature of the MHD waves.
As a plausible scenario, we thus suggest that mode 2 is excited (as a second harmonic) by a standing slow magnetoacoustic wave. Mode 1 however is produced through global (slow-mode) kink oscillations in the same loop which, by periodically perturbing the plasma density in the vicinity of the X-point of a nearby flaring loop, play the role of directly triggering the modulation of the emission of flaring energy.
In this case, the mechanism responsible for the periodically triggered reconnection could be that envisaged by \cite{Chen2006}.
This would also explain the different amplitude modulation of the two oscillations.
In fact, mode 2 appears before mode 1 and displays the typical decaying oscillatory pattern usually observed for standing slow magnetoacoustic waves excited in stellar coronal loops.
As for mode 1, its shape would critically depend on factors such as the damping of the global kink wave in the first loop and the efficiency of the acceleration mechanism in the second loop that would probably be enhanced, also for geometrical reasons, in the first few cycles of the global kink oscillation.
In view of the above discussion, we thus favor this latter interpretation as more plausible (albeit not unequivocal).

On the other hand, the binary nature of the system comprising KIC 8414845 could hint at an alternative scenario.
\cite{Doyle1990} reported photometric observations of repetitive apparently periodic ($P=48\pm3$ min) flares on the binary star YY Gem.
Supported by 2.5D MHD numerical simulations, \cite{Gao2008} later attributed such periodicity to fast-mode magnetoacoustic waves trapped in the space between the surfaces of the two stars that modulated the magnetic reconnection responsible for the flaring.
Intriguingly, this period, originally interpreted by the authors as due to filament oscillations, is similar to that of the flare periodicities detected in this work.
Unfortunately, the Kepler observations of stars are temporally but not spatially resolved. Therefore, we cannot rule out that the two QPP modes detected by the SSA technique may be caused by completely different mechanisms and that the two periodicities are independent, that is, corresponding either to different MHD modes of the flaring region or different spatial harmonics of the same mode (see discussion in \citealt{Pugh2015}).

\section{Summary}

The light curves of stellar superflares often display QPPs with characteristic periods of tens of minutes.
However, the specific physical mechanisms responsible for these QPPs are a matter of active debate. 
A better understanding of these mechanisms is important for our general understanding of stellar flares. 
In this work, for the first time we applied a data-adaptive analysis technique, SSA, for the analysis of the QPP modulation in a superflare that occurred on KIC 8414845, which is a rapidly rotating young, solar-type star observed by the Kepler mission.

Because it is not based on a prescribed choice of the basis functions, this technique was found to be particularly suitable for analyzing the nonstationary, nonlinear QPPs observed during this stellar flare. 
The SSA has revealed that the apparent anharmonic shape of the QPP pattern was due to the superposition of two well-defined and highly significant oscillatory modes, a 49 min wavelet-like component and a 86 min component, with the first mode igniting before the peak of the flare and the second just after, as if induced by the first mode.

The peculiar amplitude modulation of the two reconstructed signals is suggestive of the presence of slow-mode transverse and/or longitudinal MHD oscillations excited in a coronal loop; this is possibly induced by flaring activity, triggering the periodic release of flaring energy in a nearby loop through a mechanism of repetitive reconnection. 
Although various scenarios have been discussed and considered as plausible, the interpretation favored in this work is that the observed QPP pattern is produced by the combination of a standing slow-mode magnetoacoustic oscillation detected at the second-harmonic and a global kink oscillation of the same coronal loop periodically triggering magnetic reconnection in a nearby loop.
In this case, a viable mechanism responsible for the modulated reconnection could be that envisaged by \cite{Chen2006} and caused by the perturbation of the plasma density in the vicinity of the reconnection site in the second loop.
However, concurrent interpretations and mechanisms cannot be ruled out on the basis of the available data since Kepler observations of stars are only temporally but not spatially resolved. 
The SSA analysis of similar multi-periodic events in the context of both solar and stellar flares is currently under way with the aim of providing more insight into the mechanisms underlying their production. 

\begin{acknowledgements} 
We thank the anonymous referee for the valuable and insightful comments and suggestions, which have improved the paper significantly.
We would like to thank the Kepler team for providing the data used in this paper, obtained from the Mikulski Archive for Space Telescopes (MAST). 
Funding for the Kepler mission is provided by the NASA Science Mission directorate.
\end{acknowledgements}

\bibliographystyle{aa}
\bibliography{Mancuso_et_al_2020_arXiv}

\begin{appendix}
\section{Singular spectrum analysis}
The first step of the SSA methodology involves embedding the time series $x = x(t); t=1,...,N$ of length $N$ into a vector space of dimension $M$ by considering $M$ lagged copies $x = x(t-j); j=1,...,M$, thus forming a lag-covariance matrix starting from the given time series.
In the formulation developed by \cite{Vautard1989}, the $M \times M$ lag-covariance matrix {\bf C$_{\rm D}$} is estimated directly from the data as a Toeplitz matrix with constant diagonals using the unbiased covariance estimator with time lag $\tau$ of the form 
\vspace{-0.1cm}
\begin{equation}
c(\tau) = {1 \over N-\tau} \sum_{n=1}^{N-\tau} x(n)x(n+\tau), 
\end{equation} 
so that ${\bf (C_{\rm D})}_{ij} = c(\tau)$, where $\tau = |i-j|$ and $i,j=1,...,M$.
The covariance matrix {\bf C$_{\rm D}$} thus evaluated from the $N$ sampled points is then diagonalized.
This is achieved by eigenvalue decomposition, yielding
$\mathbf{\Lambda} = {\rm \mathbf{E}}^{\rm T} {\rm \mathbf{C}}_{\rm D}{\rm \mathbf{E}},$ where $\mathbf{\Lambda}$ is an $M \times M$ diagonal matrix with eigenvalues \{$\lambda_k; 1\leq k \leq M$\} ranked in decreasing order and $E$ contains the corresponding eigenvectors, also known as empirical orthogonal functions (EOFs). 
The eigenvalues $\lambda_k$ yield the variance of the time series in the direction specified by the corresponding eigenvectors {\bf $E_k$} with the set $\{\sqrt{\lambda_k}; 1\leq k \leq M$\} representing the {\it singular spectrum}.  
For each EOF {\bf $E_k$}, we can construct the time series of length $N'=N-M+1$ given by the $k$-th principal component (PC) of length $N'$, 

\vspace{-0.1cm}
\begin{equation}
A_k(i) = \sum_{k=1}^M x(i+j-1)E_k(j), \qquad 1\leq i \leq N',
\end{equation}
which represent the projection of the original time series onto the $k$-th EOF. 
In order to recover the phase information, the PCs are then combined to form partial reconstructions (RCs) of length $N$ of the original time series as follows:
\vspace{-0.1cm}
\begin{equation}
R_k(t) =  {1 \over M_t}  \sum_{k\in K} \sum_{j=L_t}^{U_t} A_k(t-j+1)E_k{j}, \qquad 1\leq t \leq N,
\end{equation}
where $(M_t, L_t, U_t)$ corresponds to $(t,1,t)$ for $1 \leq t \leq M-1$, $(M,1,M)$ for $M \leq t \leq N'$, and $(N-t+1,t-N+M,M)$ for $N'+1\leq t \leq N$.
The time series can be reconstructed completely by summing all RCs.

\end{appendix}

\end{document}